# Securing OFDM-Based Wireless Links Using Temporal Artificial-Noise Injection

Mohamed F. Marzban[†], Ahmed El Shafie[†], Rakan Chabaan[⋆], Naofal Al-Dhahir[†]

[†]University of Texas at Dallas, USA.
[⋆]Hyundai-Kia America Tech. Center, Inc.

*Abstract*—We investigate the physical layer security of wireless single-input single-output orthogonal-division multiplexing (OFDM) when a transmitter, which we refer to as Alice, sends her information to a receiver, which we refer to as Bob, in the presence of an eavesdropping node, Eve. To prevent information leakage, Alice sends an artificial-noise (AN) signal superimposed over her information signal. We investigate the impact of the channel delay spread, OFDM cyclic prefix, information/AN power allocation, and information and AN precoders design on the achievable average secrecy rate. We consider the two cases of known and unknown channel state information (CSI) at Alice. Furthermore, we compare both cases of per-sub-channel processing and joint sub-channels processing at Eve's receiver. Our numerical results show the gains of AN injection in terms of average secrecy rate for different OFDM operating conditions. Moreover, based on our new insights, we demonstrate that the AN-aided scheme is effective and achieves almost the same average secrecy rate as the full-CSI case without the need for Eve's instantaneous CSI at Alice.

*Index Terms*—OFDM, artificial noise, security, wiretap channel

## I. INTRODUCTION

Information transmitted over the wireless channel is vulnerable to eavesdropping due to the broadcast nature of the wireless medium. A wireless node with an off-the-shelf radio module can send/receive data as well as freely listen to other communications within radio range. This is a critical issue since an eavesdropping node can analyze the identity of the transmitting terminals. In addition, an eavesdropping node can obtain important information regarding the activity and data shared between the legitimate nodes. Security has been always maintained through encryption algorithms deployed over the upper layers of the protocol stack. However, such algorithms assume limited computation power at the eavesdropping node. The secrecy of communications between two legitimate parties in an information-theoretic sense, known as the physical-layer (PHY) security, was first investigated in the seminal work of Wyner [1]. The system's PHY security is measured by the secrecy capacity of the link connecting the legitimate parties, which represents the maximum rate of the legitimate parties with zero information leakage at the eavesdropping node. PHY security approaches exploit the randomness of the channel in securing the data between the communicating nodes. This is achievable by sacrificing a portion of Bob's transmission rate in order to confuse nearby eavesdroppers [2].

Investigating the security of OFDM-based systems is crucial since most of the wired and wireless communication systems adopt OFDM modulation/demodulation. The secrecy capacity of OFDM-based systems were derived in [3]. To simplify the system's design, the authors of [4] defined the system's security using a lower-bound on the minimum mean squared error between the transmitted and decoded data at the eavesdropper. By modeling the OFDM wiretap channel as a special instance of the multiple-input multiple-output (MIMO) wiretap channel, the authors of [5] derived its secrecy rates for asymptotically high and low signal-to-noise ratio (SNR) regimes.

To further confuse the eavesdroppers, linear precoding for PHY security was proposed in, e.g., [6]–[11] and the references therein. For example, in [8], the authors proposed power-allocation schemes for precoding transmissions aided with artificial noise (AN). Reference [9] investigated temporal-AN injection scheme for the single-input single-output single-antenna eavesdropper (SISOSE) OFDM system. The temporal degrees of freedom provided by adding the cyclic-prefix sequence is exploited to transmit a precoded time-domain AN vector which is added to the information vector before transmission. A direct extension of the problem presented to the MIMO case was proposed in [10]. In [11], the authors investigated the impact of a hybrid spatial-temporal AN scheme on the security of MIMOME-OFDM systems. In [12], a secure scheme for combining temporal AN with secret key extraction is proposed. Different from the above-mentioned works, we investigate the properties of temporal AN and analyze its dependency on the channel matrices which follow a circulant structure. Then, we exploit these properties to design precoding and power allocation strategies for the case when Eve's instantaneous channel state information (CSI) is available at Alice and the case when Alice does not know Eve's channel. Our main contributions in this paper are as follows

- We exploit the temporal degrees of freedom provided by the cyclic prefix (CP) sequence added to each OFDM symbol to inject AN vectors that hurt the illegitimate receivers (i.e., Eves). We investigate the temporal AN scheme and study the precoders' designs at Alice that can increase the average secrecy rate.
- We study the power allocation for data and AN signals under known and unknown Eve's instantaneous CSI at Alice. We analyze and break down the AN-precoding problem into two AN precoders. The first precoding matrix cancels the AN at the legitimate receiver, Bob, while the second precoding matrix, assuming known Eve's instantaneous CSI at Alice, focuses the AN signal in the directions of the Alice-Eve channel vectors. Then, we show that, with an appropriate design of the first precoder, the second precoder can be eliminated which reduces the design complexity. Moreover, we show that

This work is supported by Hyundai Inc.

there is a useful number of directions that can be used to confuse Eve (i.e., degrade her data rate) which is a function of the delay spread duration of both the Alice-Bob and Alice-Eve channels.
- We derive a tight approximation on the instantaneous secrecy rate for the case of equal-power allocation and show that, at high input SNR, the average secrecy rate is a linear function of Alice's transmit power level (in dB). Hence, unlike the case of no AN injection where the instantaneous secrecy rate becomes independent of the transmit power level, in our investigated AN injection scheme, increasing Alice's transmit power can still increase the instantaneous secrecy rate. Moreover, the achievable average secrecy rate is a linear function of the number of useful AN streams.
- Unlike [9], we investigate the best decoding strategy at Eve, where Eve jointly decodes the OFDM sub-channels simultaneously using the maximum likelihood (ML) decoder, and we derive its instantaneous secrecy rates. Then, we compare it with the per sub-carrier processing strategy and show that Eve's data rate significantly decreases when she uses per sub-channel processing as in [9]. This fact demonstrates that for Eve to achieve the highest possible rate, she needs to adopt the complicated ML decoder whose complexity increases exponentially with the number of OFDM sub-channels.
- We propose a new transmit power allocation scheme which significantly reduces the complexity of the suboptimal scheme proposed in [9]. Our scheme achieves a comparable performance as the suboptimal iterative scheme proposed in the literature with much lower complexity and we derive closed-form expressions for the used power levels.

*Notation:* Lower- and upper-case bold letters denote vectors and matrices, respectively. $\mathbf{I}_N$ and $\mathbf{F}$ denote, respectively, the identity matrix whose size is $N \times N$ and the fast Fourier transform (FFT) matrix. $\mathbb{C}^{M \times N}$ denotes the set of all complex matrices of size $M \times N$. $(\cdot)^\top$ and $(\cdot)^*$ denote transpose and Hermitian (i.e., complex-conjugate transpose) operations, respectively. $|\cdot|$ cardinality of a set. $\mathbb{R}^{M \times N}$ denotes the set of real matrices of size $M \times N$. $\mathbb{E}\{\cdot\}$ denotes statistical expectation. $[\mathbf{A}]_{k,1:N}$ is the $k$-th row of the matrix $\mathbf{A} \in \mathbb{C}^{M \times N}$. $\mathbf{0}_{M \times N}$ denotes the all-zero matrix with size $M \times N$. $\mathrm{diag} = \{\cdot\}$ denotes a diagonal matrix with the enclosed elements as its diagonal elements. $[\cdot]^+ = \max\{0, \cdot\}$ returns the maximum between the argument and zero.

## II. SYSTEM MODEL AND TEMPORAL AN DESIGN

In this section, we state our main assumptions and then explain the design of the proposed AN-aided scheme.

### A. System Model and Assumptions

Consider a wiretap communication scenario consisting of a single transmitter (Alice) who wishes to send a confidential information to a legitimate receiver which we refer to as Bob, in the presence of an eavesdropper (Eve). All three nodes are equipped with single antennas. Alice transmits her data using OFDM with $N$ sub-channels. The CP of size $N_{\mathrm{cp}}$ is added in time domain at the transmitter using the CP insertion matrix, denoted by $\mathbf{T}^{\mathrm{cp}} = \begin{bmatrix} \mathbf{E}_{N_{\mathrm{cp}} \times N}^\top & \mathbf{I}_N \end{bmatrix}^\top \in \mathbb{R}^{(N+N_{\mathrm{cp}}) \times N}$, and is removed at the receiver using the CP removal matrix, denoted by $\mathbf{R}^{\mathrm{cp}} = \begin{bmatrix} \mathbf{0}_{N \times N_{\mathrm{cp}}} & \mathbf{I}_N \end{bmatrix} \in \mathbb{R}^{N \times (N+N_{\mathrm{cp}})}$, where $\mathbf{E} = \begin{bmatrix} \mathbf{0}_{N_{\mathrm{cp}} \times (N-N_{\mathrm{cp}})} & \mathbf{I}_{N_{\mathrm{cp}}} \end{bmatrix}$. The CP is used to eliminate the inter-block interference and should be longer than the channel delay spreads. Let $L_{\mathrm{B}}$ and $L_{\mathrm{E}}$ denote the channel memories of Alice-Bob and Alice-Eve links, respectively. We assume that the channel coefficients of Alice-Bob and Alice-Eve links remain constant during the coherence time and change from one coherence time to another. Each tap of the channel impulse response (CIR) of the Alice-Bob and Alice-Eve links has average gain of $\sigma^2_{\mathrm{A-B}}$ and $\sigma^2_{\mathrm{A-E}}$, respectively. The CP is designed to be longer than the channel memory of both the Alice-Bob and Alice-Eve links. Let $\mathbf{n}_{\mathrm{B}}$ and $\mathbf{n}_{\mathrm{E}}$ denote the zero mean circularly-symmetric complex-additive Gaussian white noise (AWGN) at Bob and Eve, respectively. Let $\kappa_{\mathrm{B}}$ and $\kappa_{\mathrm{E}}$ denote the noise variances of Bob and Eve, respectively. Hence, for a per sub-channel bandwidth of $\Delta_f$, the noise power per sub-channel at Bob and Eve are $\Delta_f \kappa_B$ and $\Delta_f \kappa_{\mathrm{E}}$, respectively.

Let $\mathbf{H}^{\mathrm{time}} \in \mathbb{C}^{(N+N_{\mathrm{cp}}) \times (N+N_{\mathrm{cp}})}$ and $\mathbf{G}^{\mathrm{time}} \in \mathbb{C}^{(N+N_{\mathrm{cp}}) \times (N+N_{\mathrm{cp}})}$ denote the complex time-domain Toeplitz channel matrices of the Alice-Bob and Alice-Eve links, respectively. Moreover, $\mathbf{H} = \mathbf{F}\mathbf{R}^{\mathrm{cp}}\mathbf{H}^{\mathrm{time}}\mathbf{T}^{\mathrm{cp}}\mathbf{F}^* \in \mathbb{C}^{N \times N}$ and $\mathbf{G} = \mathbf{F}\mathbf{R}^{\mathrm{cp}}\mathbf{G}^{\mathrm{time}}\mathbf{T}^{\mathrm{cp}}\mathbf{F}^* \in \mathbb{C}^{N \times N}$ denote the corresponding frequency-domain complex diagonal channel coefficients of Alice-Bob and Alice-Eve links, respectively. $\mathbf{P}_{\mathrm{x}} = \mathrm{diag}(p_{x_1}, p_{x_2}, \cdots, p_{x_N})$ is a diagonal power matrix containing the transmit power levels at each sub-channel, and $\mathbf{x}$ denotes the $N \times 1$ independent data symbols with zero mean and unit variance. The $N \times 1$ received signals at Bob and Eve are given respectively by,

$$\mathbf{y}_{\mathrm{B}} = \mathbf{H}\mathbf{P}_{\mathrm{x}}^{\frac{1}{2}}\mathbf{x} + \mathbf{n}_{\mathrm{B}}, \quad \text{and} \quad \mathbf{y}_{\mathrm{E}} = \mathbf{G}\mathbf{P}_{\mathrm{x}}^{\frac{1}{2}}\mathbf{x} + \mathbf{n}_{\mathrm{E}} \quad (1)$$

### B. Temporal AN Design

The temporal AN enhances the PHY security by sacrificing a portion of Alice's data power to transmit AN signals to degrade Eve's SNR. The AN is injected in the time domain to exploit the temporal degrees of freedom provided by the CP [9]. Without loss of optimality, we assume that the AN precoding is composed of two matrices. The first matrix, $\mathbf{Q} \in \mathbb{C}^{(N+N_{\mathrm{cp}}) \times N_{\mathrm{cp}}}$, designs the AN to span the null space of the channel matrix between Alice and Bob as follows

$$\mathbf{R}^{\mathrm{cp}}\mathbf{H}^{\mathrm{time}}\mathbf{Q} = 0 \quad (2)$$

The second precoding matrix, $\mathbf{U} \in \mathbb{C}^{N_{\mathrm{cp}} \times N_s}$, correlates the AN streams and steers them to maximize the interference power at Eve. Given that $\mathbf{Q}$ has $N_{\mathrm{cp}}$ orthonormal basis vectors, we can send a maximum of $N_{\mathrm{cp}}$ AN streams. Let $\mathbf{z} \sim \mathcal{CN}(0, \mathbf{\Sigma}_z)$ denote the $N_s \times 1$ AN symbols transmitted to confuse Eve with $N_s \leq N_{\mathrm{cp}}$ streams and $\mathbf{\Sigma}_z$ denote the AN diagonal covariance matrix. The received signals at Bob and Eve can be, respectively, expressed as

$$\mathbf{y}_{\mathrm{B}} = \mathbf{H}\mathbf{P}_{\mathrm{x}}^{\frac{1}{2}}\mathbf{x} + \mathbf{n}_{\mathrm{B}} \quad (3)$$

$$\mathbf{y}_{\mathrm{E}} = \mathbf{GP}_{\mathrm{x}}^{\frac{1}{2}}\mathbf{x} + \mathbf{FR}^{\mathrm{cp}}\mathbf{G}^{\mathrm{time}}\mathbf{QUz} + \mathbf{n}_{\mathrm{E}} \qquad (4)$$

**Proposition 1.** *Let $L_{\mathrm{u}}$ denote the maximum of the channel memories of the Alice-Bob and Alice-Eve links (i.e., $L_{\mathrm{u}} = \max(L_{\mathrm{B}}, L_{\mathrm{E}})$). There are $(N_{\mathrm{cp}} - L_{\mathrm{u}})$ useless AN directions that project the AN signals in the null space of Alice-Eve channel, and $N_{\mathrm{s}} = L_{\mathrm{u}}$ useful AN directions that can degrade Alice-Eve's instantaneous link rate.*

*Proof.* Alice superimposes the temporal AN symbols on the data signal in the time domain after CP insertion. The composite signal is received at Bob after it passes through the equivalent Alice-Bob channel in the time domain which is expressed as $\mathbf{R}^{\mathrm{cp}}\mathbf{H}^{\mathrm{time}}$, where $\mathbf{R}^{\mathrm{cp}} = [\mathbf{0}_{N \times N_{\mathrm{cp}}} \mathbf{I}_N]$ and $\mathbf{H}^{\mathrm{time}}$ is lower-triangular Toeplitz channel matrix with $[h^{\mathrm{time}}(0), h^{\mathrm{time}}(1), \ldots, h^{\mathrm{time}}(L_{\mathrm{B}}), 0, \ldots, 0]^\top$ as its first column where $h^{\mathrm{time}}(i)$ is the $i$-th tap of the Alice-Bob link CIR. Hence, $\mathbf{R}^{\mathrm{cp}}\mathbf{H}^{\mathrm{time}}$ can be expressed as

$$\mathbf{R}^{\mathrm{cp}}\mathbf{H}^{\mathrm{time}} = \begin{bmatrix} \mathbf{0}_{N \times (N_{\mathrm{cp}} - L_{\mathrm{B}})} & \mathbf{H}'_{N \times (N + L_{\mathrm{B}})} \end{bmatrix} \qquad (5)$$

where $\mathbf{H}' \in \mathbb{C}^{N \times N}$ is the equivalent Alice-Bob channel matrix after CP removal which is an upper-triangular Toeplitz matrix with $[h^{\mathrm{time}}(L_{\mathrm{B}}), h^{\mathrm{time}}(L_{\mathrm{B}} - 1), \ldots, h^{\mathrm{time}}(0), 0 \ldots, 0]$ as its first row. The null space precoder ($\mathbf{Q}$) contains the orthonormal null space basis vectors of the equivalent Alice-Bob channel matrix which can be expressed as

$$\mathbf{Q} = \mathrm{Null}\left(\mathbf{R}^{\mathrm{cp}}\mathbf{H}^{\mathrm{time}}\right) = \mathrm{Null}\left(\begin{bmatrix} \mathbf{0}_{N \times (N_{\mathrm{cp}} - L_{\mathrm{B}})} & \mathbf{H}'_{N \times (N + L_{\mathrm{B}})} \end{bmatrix}\right). \qquad (6)$$

$\mathbf{R}^{\mathrm{cp}}\mathbf{H}^{\mathrm{time}}$ has $N_{\mathrm{cp}} - L_{\mathrm{B}}$ zero columns. Selecting any of these all-zero columns has to be part of the orthonormal basis of the Alice-Bob channel matrix null space which is given by

$$\mathbf{Q} = \mathrm{Null}\left(\mathbf{R}^{\mathrm{cp}}\mathbf{H}^{\mathrm{time}}\right) = \begin{bmatrix} \mathbf{W}_{(N+N_{\mathrm{cp}}) \times (N_{\mathrm{cp}} - L_{\mathrm{B}})} & \mathbf{Q}'_{(N+N_{\mathrm{cp}}) \times L_{\mathrm{B}}} \end{bmatrix} \qquad (7)$$

where $\mathbf{W} = [\mathbf{I}_{N_{\mathrm{cp}} - L_{\mathrm{B}}} \ \mathbf{0}_{(N+L_{\mathrm{B}}) \times (N_{\mathrm{cp}} - L_{\mathrm{B}})}]^\top$ and $\mathbf{Q}' = \mathrm{Null}(\mathbf{H}')$ contains the $L_{\mathrm{B}}$ columns of Bob's null space that arise from the different combinations of Bob's channel coefficients.

At Eve, the received AN across the $N$ data OFDM sub-channels after the FFT can be expressed as

$$\mathbf{i}_{N \times 1} = \tilde{\mathbf{Q}}\mathbf{z} \qquad (8)$$

where $\tilde{\mathbf{Q}} = \mathbf{FR}^{\mathrm{cp}}\mathbf{G}^{\mathrm{time}}\mathbf{Q}$ of size $N \times N_{\mathrm{cp}}$ represents the equivalent channel matrix experienced by the temporal AN before it reaches Eve. Eve's time-domain channel after CP removal is given by

$$\mathbf{R}^{\mathrm{cp}}\mathbf{G}^{\mathrm{time}} = \begin{bmatrix} \mathbf{0}_{N \times (N_{\mathrm{cp}} - L_{\mathrm{E}})} & \mathbf{G}'_{N \times (N+L_{\mathrm{E}})} \end{bmatrix} \qquad (9)$$

Hence, the AN vector across the data OFDM sub-channels of Eve can be expressed as

$$\mathbf{i}_{N \times 1} = \mathbf{F}_{N \times N} \begin{bmatrix} \mathbf{0}_{N \times (N_{\mathrm{cp}} - L_{\mathrm{E}})} & \mathbf{G}'_{N \times (N+L_{\mathrm{E}})} \end{bmatrix} \\ \times \begin{bmatrix} \mathbf{W}_{(N+N_{\mathrm{cp}}) \times (N_{\mathrm{cp}} - L_{\mathrm{B}})} & \mathbf{Q}'_{(N+N_{\mathrm{cp}}) \times L_{\mathrm{B}}} \end{bmatrix} \mathbf{z}_{N_{\mathrm{cp}} \times 1} \qquad (10)$$

where $\mathbf{G}'$ represents the non-zero columns of the equivalent time-domain Alice-Eve channel matrix after CP removal which has a form similar to $\mathbf{H}'$. The number of zero columns that project AN on the CP of Eve is equal to $\min(N_{\mathrm{cp}} - L_{\mathrm{B}}, N_{\mathrm{cp}} - L_{\mathrm{E}}) = N_{\mathrm{cp}} - \max(L_{\mathrm{B}}, L_{\mathrm{E}})$. Hence, the equivalent AN at Eve is given by

$$\mathbf{i}_{N \times 1} = \mathbf{F}_{N \times N}[\mathbf{0}_{N \times (N_{\mathrm{cp}} - L_{\mathrm{u}})} \ \mathbf{U}'_{\mathrm{useful} \ N \times L_{\mathrm{u}}}] \mathbf{z}_{N_{\mathrm{cp}} \times 1} \qquad (11)$$

where $\mathbf{U}'_{\mathrm{useful}} = \begin{bmatrix} \mathbf{0}_{N \times (N_{\mathrm{cp}} - L_{\mathrm{E}})} & \mathbf{G}'_{N \times (N+L_{\mathrm{E}})} \end{bmatrix} \mathbf{Q}'$, and $L_{\mathrm{u}} = \max(L_{\mathrm{B}}, L_{\mathrm{E}})$. From Equation (11), we can deduce that $\tilde{\mathbf{Q}}$ has rank $L_{\mathrm{u}}$. Furthermore, we conclude that the first $N_{\mathrm{cp}} - L_{\mathrm{u}}$ AN streams of $\mathbf{z}_{N_{\mathrm{cp}} \times 1}$ are useless and do not impact Eve's performance since they are multiplied by all-zero columns. These useless streams inject AN only on the CP of Eve which will be removed anyway by the receiver before processing the data streams. The useful streams are the $L_{\mathrm{u}}$ AN streams multiplied by $\mathbf{U}'_{\mathrm{useful}}$. Hence, the number of useful AN streams is equal to the maximum of Alice-Bob and Alice-Eve channel memories. As $L_{\mathrm{B}}$ increases, the number of non-zero orthonormal basis vectors of Bob's null space increases. Therefore, the number of useful AN streams increases. As $L_{\mathrm{E}}$ increases, Eve's channel spreads each AN stream across more of its data sub-channels. Hence, the number of useful streams is $L_{\mathrm{u}} = \max(L_{\mathrm{B}}, L_{\mathrm{E}})$. □

In the following subsections, we analyze the two cases where Eve's instantaneous CSI is known or unknown at Alice which affect the design of the second precoder matrix, $\mathbf{U}$.

### C. Eve's Instantaneous CSI is Known at Alice

In this subsection, we exploit the availability of Eve's instantaneous CSI at Alice to design a second precoder ($\mathbf{U}$) to maximize the interference at Eve and extract the $L_{\mathrm{u}}$ useful streams. Alice-Eve's instantaneous link rate can be expressed as

$$R_{\mathrm{E}} = \log_2 \det\left(\mathbf{I}_N + \mathbf{GP}_{\mathbf{x}}\mathbf{G}^*\left(\tilde{\mathbf{Q}}\mathbf{U}\boldsymbol{\Sigma}_z\mathbf{U}^*\tilde{\mathbf{Q}}^* + \Delta_f\kappa_{\mathrm{E}}\mathbf{I}_N\right)^{-1}\right) \qquad (12)$$

From Proposition 1, $\tilde{\mathbf{Q}}$ has rank $L_{\mathrm{u}}$. Hence, it has $L_{\mathrm{u}}$ non-zero singular values and $N_{\mathrm{cp}} - L_{\mathrm{u}}$ zero singular values. Considering the singular value decomposition (SVD) $\tilde{\mathbf{Q}} = \mathbf{R}_z\boldsymbol{\Lambda}_z\mathbf{V}_z^*$, Alice-Eve's instantaneous rate is given by

$$R_{\mathrm{E}} = \log_2 \det(\mathbf{I}_N + \mathbf{P}_{\mathbf{x}}\mathbf{GG}^* \\ \times (\mathbf{R}_z\boldsymbol{\Lambda}_z\mathbf{V}_z^*\mathbf{U}\boldsymbol{\Sigma}_z\mathbf{U}^*\mathbf{V}_z\boldsymbol{\Lambda}_z^*\mathbf{R}_z^* + \Delta_f\kappa_{\mathrm{E}}\mathbf{I}_N)^{-1}) \qquad (13)$$

To maximize the interference power at Eve, we concentrate all AN power on the useful $L_{\mathrm{u}}$ streams. We use $N_s = N_{\mathrm{cp}}$ and assign zero power to the useless directions. We design $\mathbf{U}$ such that $\mathbf{U}^*\mathbf{V}_z = \mathbf{I}_{N_{\mathrm{cp}}}$ and steer the AN to the right singular vectors of Eve's channel matrix that correspond to the non-zero singular values. Doing so, we extract the $L_{\mathrm{u}}$ useful AN streams, and Alice-Eve's link instantaneous rate is given by

$$R_{\mathrm{E}} = \log_2 \det\left(\mathbf{I}_N + \mathbf{P}_{\mathbf{x}}\mathbf{GG}^*\left(\mathbf{R}_z\boldsymbol{\Lambda}_z\boldsymbol{\Sigma}_z\boldsymbol{\Lambda}_z^*\mathbf{R}_z^* + \Delta_f\kappa_{\mathrm{E}}\mathbf{I}_N\right)^{-1}\right). \qquad (14)$$

The instantaneous rate of the Alice-Bob link is given by

$$R_{\mathrm{B}} = \sum_{k=1}^{N} \log_2\left(1 + \frac{|H_k|^2 p_{x,k}}{\Delta_f \kappa_{\mathrm{B}}}\right) \qquad (15)$$

where $H_k$ is the channel coefficient at OFDM sub-channel $k$. We emphasize here that the Alice-Bob channel matrix is diagonal and the noise is AWGN and, hence, the per sub-channel processing at Bob is optimal. On the contrary, the temporal AN couples the sub-channels at Eve's receivers. Hence, she needs to perform the joint sub-channels processing/decoding to achieve the highest possible data rate in (14). The instantaneous secrecy rate of the legitimate system is given by

$$R_{\mathrm{sec}} = [R_{\mathrm{B}} - R_{\mathrm{E}}]^+ \qquad (16)$$

The power allocation problem is formulated to maximize $R_{\mathrm{sec}}$ subject to a total power constraint as follows

$$\max_{\mathbf{P}_{\mathbf{x}}, \boldsymbol{\Sigma}_z}: \sum_{k=1}^{N} \log_2\left(1 + \frac{|H_k|^2 p_{x,k}}{\Delta_f \kappa_{\mathrm{B}}}\right) \\ - \log_2 \det\left(\mathbf{I}_N + \mathbf{P}_{\mathbf{x}}\mathbf{GG}^*\left(\mathbf{R}_z\boldsymbol{\Lambda}_z\boldsymbol{\Sigma}_z\boldsymbol{\Lambda}_z^*\mathbf{R}_z^* + \Delta_f\kappa_{\mathrm{E}}\mathbf{I}_N\right)^{-1}\right) \qquad (17)$$

$$\text{s.t.} \sum_{i=1}^{N} p_{x,i} + \sum_{i=1}^{L_u} p_{z,i} = P \quad (a)$$
$$p_{x,i} \geq 0, \quad \forall i \in \{1, 2, \ldots, N\} \quad (b)$$
$$p_{z,i} \geq 0 \quad \forall i \in \{1, 2, \ldots, L_u\} \quad (c)$$

where $p_{z,i}$ are the diagonal elements of the diagonal covariance matrix $\Sigma_z$ corresponding to the non-zero singular values, and $P$ denotes the total power of data and AN. Problem (17) is non-convex in the optimization variables $(p_{x,i}, p_{z,i})$. Hence, we propose a suboptimal two stage approach, where in the first stage, we obtain the power allocation for the AN streams and in the second stage, the power assignments to the data sub-channels are computed.

In the first stage, we treat the AN symbols as data symbols that should be received at Eve under the rate-maximization problem. In other words, the AN power allocation problem is formulated as follows

$$\max_{p_{z,i}}: \log_2 \det(\mathbf{I}_N + \mathbf{R}_z \mathbf{\Lambda}_z \mathbf{\Sigma}_z \mathbf{\Lambda}_z^* \mathbf{R}_z^*), \quad \text{s.t.} \sum_{i=1}^{L_u} p_{z,i} = (1-\alpha)P \quad (18)$$

where $\alpha$ is the fraction of power allocated to data and $(1-\alpha)$ is the power fraction dedicated to AN. By using Sylvester's identity, Problem (18) can be reformulated as follows

$$\max_{p_{z,i}}: \sum_{i=1}^{L_u} \log_2(1 + \delta_{z,i}^2 p_{z,i}), \quad \text{s.t.} \sum_{i=1}^{L_u} p_{z,i} = (1-\alpha)P \quad (19)$$

where $\delta_{z,i}$ denotes the $i$-th diagonal element of the diagonal matrix $\mathbf{\Lambda}_z$. Problem (19) is a convex optimization problem and can be solved using the well-known water filling approach as follows

$$p_{z,i} = \left[C_1 - \frac{1}{\delta_{z,i}^2}\right]^+ \quad \forall\, i \in 1, 2, \cdots, L_u \quad (20)$$

where $C_1$ is a constant and can be obtained as follows

$$C_1 = \frac{(1-\alpha)P}{L_u} + \frac{1}{L_u} \sum_{k=1}^{L_u} \frac{1}{\delta_{z,i}^2} \quad (21)$$

After obtaining the power allocation for the AN symbols, in the second stage, the data power allocation can be obtained from [9], [13] as follows

$$p_i = \begin{cases} \frac{1}{2\xi_i}\left(\sqrt{\gamma_i^2 - 4\xi_i \frac{\mu + |\tilde{G}_i|^2 - |H_i|^2}{\mu}} - \gamma_i\right), & |H_i|^2 - |\tilde{G}_i|^2 > \mu \\ 0, & \text{otherwise} \end{cases} \quad (22)$$

where $\gamma_i = |H_i|^2 + |\tilde{G}_i|^2$, $\xi_i = |H_i|^2|\tilde{G}_i|^2$, $|\tilde{G}_i|^2 = \frac{|G_i|^2}{\Delta_f \kappa_E + q_i^* U \Sigma_z U^* q_i}$, $q_i$ represents the $i$-th row of $\tilde{\mathbf{Q}}$, and $\mu$ denotes the Lagrangian multiplier which can be obtained form the total power constraint $\sum_i^N p_i = \alpha P$.

### D. Eve's CSI is Unknown at Alice

When Alice does not know the coefficients of Alice-Eve's channel (represented here by $\mathbf{U}'_{\text{useful}}$), she can not use a second precoder ($\mathbf{U}$). In this case, the instantaneous Alice-Eve's link rate can be expressed as follows,

$$R_E = \log_2 \det \left(\mathbf{I}_N + \mathbf{GP_x G}^* \left(\tilde{\mathbf{Q}}\Sigma_z \tilde{\mathbf{Q}}^* + \Delta_f \kappa_E \mathbf{I}_N\right)^{-1}\right) \quad (23)$$

However, Alice can still extract the $L_u$ useful streams that can actually harm Eve with only the knowledge of Alice-Eve's channel memory ($L_E$). This can be achieved by designing the null space precoder as in Equation (7) while the AN covariance matrix is designed as follows

$$\mathbf{\Sigma}_z = \text{diag}(\underbrace{0, 0, \cdots 0}_{N_{\text{cp}}-L_u \text{ terms}}, \underbrace{p_z, p_z, \cdots, p_z}_{L_u \text{ terms}}) \quad (24)$$

By substituting with Equations (7) and (24) into Eve's instantaneous rate in (23), we can notice that no power is allocated to the useless AN streams. Equal power allocation is performed across the useful AN streams since Alice does not know which streams cause more harm to Eve. The power of each AN stream is given by $p_z = \frac{(1-\alpha)P}{L_u}$. Without Eve's channel knowledge, the data power is allocated only based on the CSI of the Alice-Bob link. The data power allocation problem can be formulated as follows

$$\max_{p_{x,k}}: \sum_{k=1}^{N} \log_2 \left(1 + \frac{p_{x,k}|H_k|^2}{\Delta_f \kappa_B}\right), \quad \text{s.t.} \sum_{k=1}^{N} p_{x,k} = \alpha P \quad (25)$$

Problem (27) is the well-known water filling problem where the optimal power allocation strategy is given by

$$p_{x,k} = \left[C - \frac{\Delta_f \kappa_B}{|H_k|^2}\right]^+, \quad \forall k \in \{1, 2, \cdots, N\} \quad (26)$$

where $C$ is a constant given by

$$C = \frac{\alpha P}{N} + \frac{1}{N}\sum_{k=1}^{N} \frac{\Delta_f \kappa_B}{|H_k|^2} \quad (27)$$

An approximation that we found to be very tight is to replace the unitary matrix $\mathbf{R}_z$ in Eve's instantaneous rate with the identity matrix. In this case, the instantaneous rate of the Alice-Eve link is given by

$$R_E \approx \log_2 \det\left(\mathbf{I}_N + p_x \mathbf{G}\mathbf{G}^* \mathbf{V}^{-1}\right) \quad (28)$$

where $\mathbf{V} = (\mathbf{\Lambda}_z \mathbf{\Sigma}_z \mathbf{\Lambda}_z^* + \Delta_f \kappa_E \mathbf{I}_N)$. This can be rewritten as the sum of two terms as follows

$$R_E \approx \underbrace{\sum_{i=1}^{L_u} \log_2\left(1 + \frac{|G_i|^2 p_x}{\sigma_i^2}\right)}_{\text{AWGN+AN}} + \underbrace{\sum_{i=L_u+1}^{N} \log_2\left(1 + \frac{|G_i|^2 p_x}{\Delta_f \kappa_E}\right)}_{\text{AWGN only}} \quad (29)$$

where $\sigma_i^2 = \Delta_f \kappa_E + (1-\alpha)\frac{P}{L_u}\delta_{z,i}^2$. This approximation suggests that the effect of temporal AN is equivalent to interfering on only $L_u$ of Eve's OFDM sub-channels. Hence, the instantaneous secrecy rate can be given by

$$R_{\text{sec}} \approx \left[\sum_{i=1}^{N} \log_2\left(1 + \frac{|H_i|^2 p_x}{\Delta_f \kappa_B}\right) - R_E\right]^+ \quad (30)$$

We conclude this section by illustrating the tightness of our approximation of Eve's rate in Fig. 1, which shows the average Eve's and Bob's rates and the average secrecy rate versus $\alpha$. As shown in Fig. 1 and suggested analytically by Preposition 2 the average secrecy rate becomes independent of $0 < \alpha < 1$ at high input SNR.

**Proposition 2.** *At very high input SNR, the average secrecy rate is lower bounded as follows*

$$\mathbb{E}\left\{[R_B - R_E]^+\right\} \gtrapprox L_u \log_2\left(\frac{P}{N\Delta_f \kappa_B}\right) \quad (31)$$

*Proof.* The proof is omitted here due to space limitations. $\square$

### E. Per Sub-channel Processing at Eve

Under per sub-channel processing, Eve decodes each OFDM sub-channel individually. Hence, the received signal

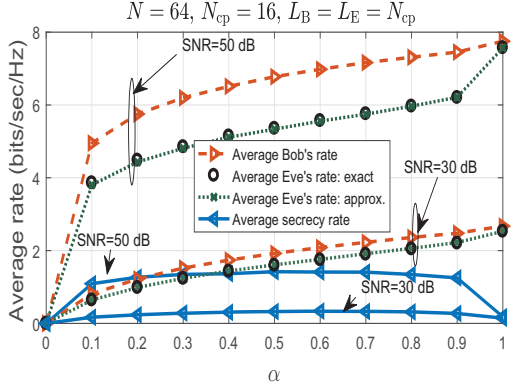

Fig. 1: Average rates and secrecy rate versus $\alpha$.

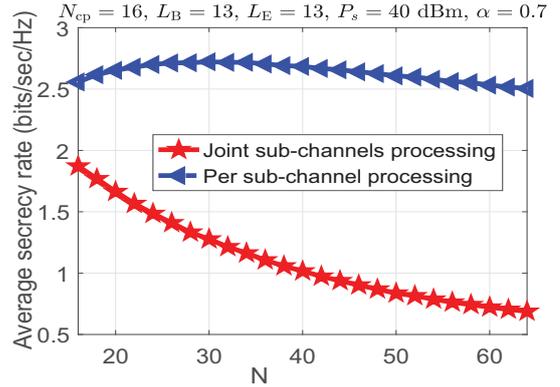

Fig. 2: Comparison between the average secrecy rate performance of per sub-channel decoding and joint OFDM sub-channels decoding at Eve.

at Eve per sub-channel $k$ is given by

$$y_{E,k} = \mathbf{G}_k\sqrt{p_{x,k}}x_k + [\mathbf{F}\mathbf{R}^{\text{cp}}\mathbf{G}^{\text{time}}\mathbf{Q}\mathbf{U}]_{k,1:N_s}\mathbf{z} + n_{E,k} \quad (32)$$

where $y_{E,k}$ is the $k$-th element of $\mathbf{y}_E$, $[\mathbf{F}\mathbf{R}^{\text{cp}}\mathbf{G}^{\text{time}}\mathbf{Q}\mathbf{U}]_{k,1:N_s}$ is the $k$-th row of $\mathbf{F}\mathbf{R}^{\text{cp}}\mathbf{G}^{\text{time}}\mathbf{Q}\mathbf{U}$, the power of the AN across the $k$-th data sub-channel is $\eta_k = [\text{diag}\{\mathbf{R}_z\boldsymbol{\Lambda}_z\boldsymbol{\Sigma}_z\boldsymbol{\Lambda}_z^*\mathbf{R}_z^*\}]_k = \sum_{i=1}^{L_u}|r_{i,k}|^2\delta_{z,i}^2 p_{z,i}$ with $\mathbf{r}_i = [r_{i,1}, r_{i,2}, \cdots, r_{i,N}]$ as the $i$-th row of $\mathbf{R}^*$, and $n_{E,k}$ is the $k$-th element of $\mathbf{n}_E$. The achievable rate at Eve under per sub-channel processing is given by

$$R_E = \sum_{i=1}^{N} \log_2\left(1 + \frac{|G_i|^2 p_{x,i}}{\eta_i + \Delta_f \kappa_E}\right) \quad (33)$$

In this case, all Eve's OFDM data sub-channels will be degraded by the AN signal which degrades Eve's instantaneous rate significantly. The achievable secrecy rate is thus given by

$$R_{\text{sec}} = \left[\sum_{i=1}^{N}\log_2\left(1+\frac{|H_i|^2 p_{x,i}}{\Delta_f\kappa_B}\right) - \sum_{i=1}^{N}\log_2\left(1+\frac{|G_i|^2 p_{x,i}}{\eta_i+\Delta_f\kappa_E}\right)\right]^+ \quad (34)$$

In contrast, the joint sub-channels processing exploits the AN correlation across the data sub-channels to minimize the interference effect to be equivalent to interference on $L_u$ sub-channels only as shown in (29).

### III. PERFORMANCE EVALUATION AND DISCUSSION

In this section, we evaluate the average secrecy rate of the system for different system's parameters. We consider a bandwidth ($B$) of one MHz and we plot the rates in bits/sec. Thus, we multiply all rate-expressions by $\frac{B}{N+N_{\text{cp}}}$ to convert the rate unit from bits/OFDM block to bits/sec. The channel taps of Bob and Eve are uniformly distributed with a total unity channel power. We assume that Bob and Eve have unit noise power (i.e. $\Delta_f\kappa_E = \Delta_f\kappa_B = 1$ dBm) and the power per sub-channel is given by $P_s = P/N$. Fig. 2 depicts the average secrecy rates versus the number of OFDM sub-channels, $N$, when Eve performs the conventional (low-complexity) per sub-channel decoding and when she performs the computationally-prohibitive joint sub-channels decoding approaches. For the joint sub-channels decoding approach, Eve can exploit the interference correlation across the data OFDM sub-channels, due to the injected temporal AN signal, and reduce its effect. We assume that equal power allocation is performed across all sub-channels. We notice that the average secrecy rate is reduced significantly when Eve performs joint OFDM sub-channels decoding and this reduction increases with $N$. As $N$ increases, the AN correlation across the data OFDM sub-channels increases and Eve can exploit such correlation to enhance her ability to decode the information. In all the following results, we assume the worst-case scenario for Alice and Bob where Eve performs joint sub-channels decoding.

Figs. 3 and 4 depict the average secrecy rate for different power-allocation algorithms at input SNR=30 dB and SNR=10 dB per sub-channel, respectively. By increasing $N_{\text{cp}}$, we increase the number of orthonormal null space vectors of the channel matrix of the Alice-Bob link. Hence, Alice can send more AN streams. However, as per Proposition 1, the number of useful AN streams only depends on the channel memories of the legitimate and eavesdropping links. Increasing $N_{\text{cp}}$ while $L_u$ is fixed only increases the number of useless AN streams which does not degrade Eve's channel. However, as $N_{\text{cp}}$ increases, the average secrecy rate decreases due to the pre-log term $\left(\frac{1}{N+N_{\text{cp}}}\right)$ in the rate expressions. Since the equal-power allocation scheme is unable to extract the useful AN streams, it has the lowest average secrecy rate. When Eve's instantaneous CSI is unknown at Alice, we can extract the useful AN streams using the properties of the null space precoder and allocate the power based on CSI of the Alice-Bob link only. The performance in this case outperforms the performance of using a second precoder with equal-power allocation. If Alice knows Eve's instantaneous CSI, she can perform the power allocation based on the per-sub-channel instantaneous secrecy rate as in (22) with a second precoder or using the null space precoder only. Both techniques have very close performance. The results are also compared with a high-complexity iterative optimization over CVX that has been proposed in [9] and similar performance is achieved with our low-complexity closed-form based scheme proposed in Section II-C. Hence, there is no need for the second precoder (i.e., precoding matrix $\mathbf{U}$) if we extracted the useful AN streams using the well-designed first precoder (i.e., precoding matrix $\mathbf{Q}$) which reduces the complexity significantly.

Fig. 5 shows the average secrecy rate versus $L_B$ for a fixed $L_E = 8$. As expected, the variation in the average secrecy rate is very small as $L_B \leq 8$. However, for $L_B > 8$, the average secrecy rate increases significantly as $L_B$ increases. Similarly, we can plot a figure for the average secrecy rate versus $L_E$ which shows similar behavior as in Fig. 5. The (average) secrecy rate depends mainly on $\max(L_B, L_E)$. Moreover Fig. 5 shows that the loss in performance, due the unavailability of Eve's CSI, is not significant since we are able to extract the useful AN streams in both scenarios.

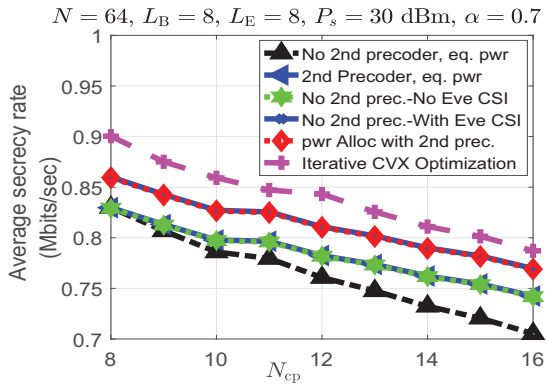

Fig. 3: Average secrecy rate for different power-allocation algorithms at SNR=30 dB as we vary $N_{cp}$.

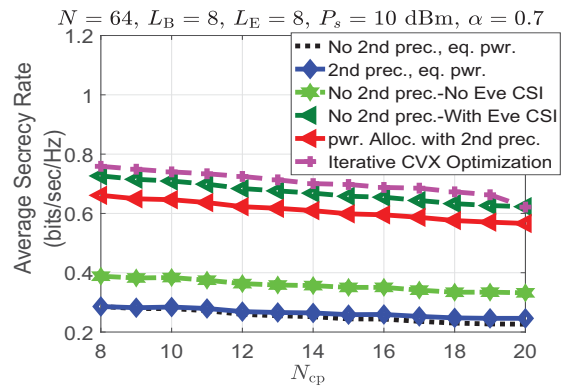

Fig. 4: Average secrecy rate for different power-allocation algorithms at SNR=10 dB as we vary $N_{cp}$.

## IV. CONCLUSIONS

We investigated the PHY security of temporal AN-aided OFDM-based wiretap communications. The main conclusions of this paper are summarized as follows

- We showed that although the null space dimension of the Alice-Bob channel matrix is $N_{cp}$, the number of useful directions that can hurt Eve is only $\max\{L_B, L_E\}$ due to the structures of Bob and Eve channel matrices after CP insertion in OFDM transmissions.
- We showed that a single null-space-based AN precoder (i.e., the first precoder) is sufficient and there is no need for a second precoder and, hence, the complexity will be reduced and there is no need to acquire Eve's instantaneous CSI at Alice.
- We show that, when Eve adopts per sub-channel processing, her rate will be significantly degraded due to the fact that the AN will be spread over all OFDM sub-channels. In contrast, if Eve adopts the joint OFDM sub-channels processing, the AN will affect only $L_u$ sub-channels. Hence, Eve's instantaneous rate will be much higher than the rate with per sub-channel processing.
- We showed that our new proposed suboptimal power allocation scheme under full CSI at Alice can achieve a comparable average secrecy rate as those in the literature, but with much lower complexity and we derived closed-form expressions for data and AN power levels.

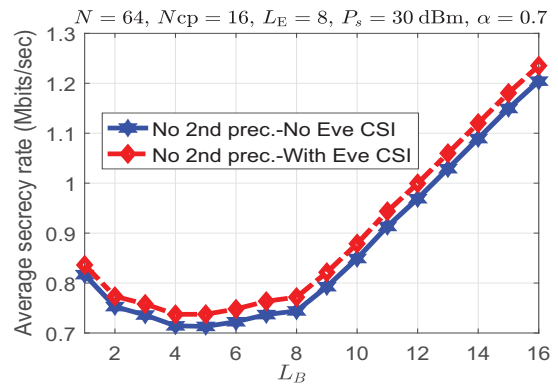

Fig. 5: Average secrecy rate versus $L_B$ for fixed $L_E$.